\documentclass{aa}  

\usepackage{graphicx}
\usepackage{placeins}
\usepackage{natbib}
\usepackage{txfonts}
\usepackage{lscape}
\usepackage{ulem}
\usepackage[colorlinks=true,citecolor=blue]{hyperref}
\makeatletter
\renewcommand*\aa@pageof{, page \thepage{} of \pageref*{LastPage}}
\makeatother

\newcommand{\MJup}{M$_{\mathrm{Jup}}$\xspace}

\newcommand{\teff}{$T_{\rm eff}$\xspace}

\begin{document}

   \title{Implications of the discovery of AF Lep b}
   \subtitle{The mass-luminosity relation for planets in the $\beta$ Pic Moving Group and the L-T transition for young companions and free-floating planets }

%   \subtitle{subtitle}

   \author{R. Gratton\inst{1}, M. Bonavita\inst{1,2}, D. Mesa\inst{1}, A. Zurlo\inst{3,4,5}, S. Marino\inst{6}, S. Desidera\inst{1}, V. D'Orazi\inst{1,7}, E. Rigliaco\inst{1}, V.  Squicciarini\inst{1,8}, P. H. Nogueira\inst{3,5}}
  \institute{
   \inst{1}INAF-Osservatorio Astronomico di Padova, Vicolo dell'Osservatorio 5, Padova, Italy, 35122-I \\
   \inst{2}Institute for Astronomy, University of Edinburgh Royal Observatory, Blackford Hill, EH9 3HJ, Edinburgh, UK \\
   \inst{3}Instituto de Estudios Astrof\'isicos, Facultad de Ingenier\'ia y Ciencias, Universidad Diego Portales, Av. Ej\'ercito 441, Santiago, Chile\\
   \inst{4}Escuela de Ingenier\'ia Industrial, Facultad de Ingenier\'ia y Ciencias, Universidad Diego Portales, Av. Ej\'ercito 441, Santiago, Chile\\
   \inst{5}Millennium Nucleus on Young Exoplanets and their Moons (YEMS), Santiago, Chile \\
   \inst{6}Department of Physics and Astronomy, University of Exeter, Stocker Road, Exeter, EX4 4QL, UK\\
   \inst{7}Dipartimento di Fisica, Universit\`{a} di Roma Tor Vergata, via della Ricerca Scientifica 1, 00133, Roma, Italy\\
   \inst{8}LESIA, Observatoire de Paris, Universit\'e PSL, CNRS, Sorbonne Universit\'e, Universit\'e Paris Cit\'e, 5 place Jules Janssen, 92195 Meudon, France
%    \inst{3}LESIA, Observatoire de Paris, Universit\'e PSL, CNRS, Sorbonne Universit\'e, Universit\'e Paris Cit\'e, 5 place Jules Janssen, 92195 Meudon, France \\
%    \inst{4}Dipartimento di Fisica, Universit\`{a} di Roma Tor Vergata, Via della ricerca scientifica, 1 - 00133, Roma, Italy \\
%    \inst{5}School of Physics and Astronomy, University of Exeter, Stocker Road, Exeter, EX4 4QL, UK \\
%   \inst{6}Instituto de Estudios Astrof\'isicos, Facultad de Ingenier\'ia y Ciencias, Universidad Diego Portales, Av. Ej\'ercito 441, Santiago, Chile \\
%  \inst{7}Escuela de Ingenier\'ia Industrial, Facultad de Ingenier\'ia y Ciencias, Universidad Diego Portales, Av. Ej\'ercito 441, Santiago, Chile \\
%\inst{8}Millennium Nucleus on Young Exoplanets and their Moons (YEMS) \\ 
%\inst{9} European Southern Observatory, Vitacura, Casilla 19001, Santiago de Chile, Chile
}

   \date{Received; accepted }

% \abstract{}{}{}{}{} 
% 5 {} token are mandatory
 
  \abstract
  % context heading (optional)
  % {} leave it empty if necessary  
   {Dynamical masses of young planets aged between 10 and 200 Myr detected in imaging play a crucial role in shaping models of giant planet formation. Regrettably, only a few such objects possess these characteristics. Furthermore, the evolutionary pattern of young sub-stellar companions in near-infrared colour-magnitude diagrams might diverge from free-floating objects, possibly due to differing formation processes. }
  % aims heading (mandatory)
   {The recent identification of a giant planet around AF Lep, part of the $\beta$~Pic moving group (BPMG), encouraged us to re-examine these points.}
  % methods heading (mandatory)
   {We considered updated dynamical masses and luminosities for the sub-stellar objects in the BPMG. In addition, we compared the properties of sub-stellar companions and free-floating objects in the BPMG and other young associations remapping the positions of the objects in the colour-magnitude diagram into a dustiness-temperature plane. }
  % results heading (mandatory)
   {We found that cold-start evolutionary models do not reproduce the mass-luminosity relation for sub-stellar companions in the BPMG. This aligns rather closely with predictions from “hot start” scenarios and is consistent with recent planet formation models. We obtain rather good agreement with masses from photometry and the remapping approach compared to actual dynamical masses. We also found a strong suggestion that the near-infrared colour-magnitude diagram for young companions is different from that of free-floating objects belonging to the same young associations.  }
  % conclusions heading (optional), leave it empty if necessary 
   {If confirmed by further data, this last result would imply that cloud settling - which likely causes the transition between L and T spectral type - occurs at a lower effective temperature in young companions than in free-floating objects. This might tentatively be explained with a different chemical composition.}


\titlerunning{Implications of the discovery of AF Lep b}
\authorrunning{R. Gratton et al.}

   \maketitle
%
%-------------------------------------------------------------------

\section{Introduction}

The detection of young planets through direct imaging enables a thorough characterisation of the individual objects. In general, the magnitude and colours can be derived, and in several cases spectra have also been obtained. This allows the derivation of the surface temperature and luminosity, though uncertainties in the model atmospheres make this derivation quite uncertain (see discussion in \citealt{Marley2015}). The best cases are those of planets at a separation of a few to a few tens of astronomical units that can be detected in stars belonging to very young and nearby associations such as the $\beta$ Pic Moving Group (BPMG: age of around 20 Myr: \citealt{Miret-Roig2020,Couture2023}). Quite accurate dynamical masses can also be obtained for these objects from their orbit and the motion of the primary, detected either through space astrometry or high-precision radial velocities \citep{Samland2017, Dupuy2022, Nowak2020, Franson2023b}. This is important in order to understand the formation of these objects and to calibrate models that are highly uncertain, especially at young ages. In fact, we expect that the early evolution, and hence the luminosity, of very young planets depends on how they assemble \citep{Spiegel2012}. Namely, a debate exists about the initial entropy of the planets, related to the fact that, in the core-accretion formation scenario of gas giant planets, most of the gas accreting onto a planet is likely processed through an accretion shock \citep{Marley2007, Mordasini2017, Berardo2017}. This shock is key in setting the structure of the forming planet and thus its observable post-formation luminosity. The radiative feedback can change the thermal and chemical structure of the circum-planetary and local circumstellar disc. Depending on the initial entropy, models with high or low initial luminosities exist (the so-called hot-start and cold-start models), though recent models suggest that hot-start is a better representation of this complex phenomenon \citep{Mordasini2017, Berardo2017}. A previous analysis based on the planets of $\beta$~Pic and that of 51~Eri indeed favours a hot start model \citep{Mordasini2017}.

We also recall that several authors (see e.g. \citealt{Liu2016, Delorme2017}) have noticed that young sub-stellar objects of L-spectral type appear redder than older ones. This fact is attributed to their lower gravity (see e.g. \citealt{Baudino2015}), the presence of a higher amount of dust in their atmospheres \citep{Chabrier2000}, or both \citep{Delorme2017}. On the other hand, the luminosity of the L-T transition for free-floating objects seems quite independent of age \citep{Dupuy2012, Liu2016}, that is, it occurs at nearly the same effective temperature for objects over a large range of ages. The transition between L and T spectral type (hereinafter L-T transition) is possibly due to the settling of dust in the atmosphere but the details of the process are still not clear \citep{Burrows2006, Burgasser2007, Saumon2008, Marley2010, Charnay2018, Vos2019}, though other scenarios involving the efficiency of vertical mixing in the atmospheres have been proposed \citep{Tremblin2016}. For instance, the models by \citet{Charnay2018} predict that the luminosity of the L-T transition is very sensitive to gravity, suggesting that it should also be sensitive to age, a fact that does not agree with observational data at least for ages $<200$~Myr. On the other hand, the size of grains likely matters; larger particles more rapidly 'rain out' of the atmosphere, leading to a sudden clearing or collapse of the clouds \citep{Knapp2004}. Finally,  the observed J-band brightening across the transition could arise from decreasing cloud coverage \citep{Ackerman2001}. We notice that while gravity does not separate objects that formed in the disc around stars from free-floating objects that formed in isolation, the presence of different amounts of dust or differences in their size or distribution might depend on their composition and then on the specific formation history of the objects. It would then be important to compare the photometric properties of young planets with those of free-floating objects of similar luminosity and age, searching for any systematic difference. \citet{Liu2016} proposed that there may indeed be some difference, but data for few planets were available at the epoch.

Unfortunately, there are only very few planets with adequate data that have been discovered so far. The addition of a single new case may have significant influence in confirming/rejecting scenarios and models. The recent discovery of a planet around AF Lep \citep{Mesa2023, DeRosa2023, Franson:2023arXiv}, a star belonging to the BPMG, prompted us to review these crucial aspects of planetary science. Even more recently, \citet{Zhang2023} presented a careful examination of the implications of the spectral energy distribution for AF Lep obtained from the previous studies on the structure of the atmosphere of this planet. This leads to a reliable determination of the effective temperature of the planet and to a strong indication that its atmosphere is much more metal-rich than that of the star.  

In this paper, we examine the implications of the discovery of AF Lep b in the derivation of the mass-luminosity relation for planets in the $\beta$ Pic moving group and on the comparison between the properties of sub-stellar objects that are either star companions or free floating. This paper is organised as follows. In Section 2, we present the main parameters for AF Lep b. In Section 3, we combine the AF Lep b parameters with those of other members of the BPMG to discuss the mass-luminosity relation for 20 Myr old planets. In Section 4, we compare the colour-magnitude diagram for young planets and free-floating objects for the members of the BPMG and of other young associations. We draw conclusions in Section 5. The Appendices contain a compilation of data for sub-stellar companions and free-floating objects belonging to young and intermediate age associations used in this paper, and the derivation of a uniform set of temperature and masses for them.

\section{Parameters for AF Lep b}

\renewcommand{\arraystretch}{1.25}
\begin{table*}
\caption{Comparison of parameters for AF Lep b derived in this study with those reported in previous works.}
\centering
\label{tab:comparison}
\begin{tabular}{llllll}
\hline
\teff(K)    & $\log{g}$    & $\log{L/L_{\odot}}$   & Mass (\MJup)          & Source & Model\\
\hline
$908\pm 123$  & $-3.7\pm 0.3$  & $-4.97\pm 0.20$ & $3.2^{+0.7}_{-0.6}$    & this study & AMES-COND \\
1000-1700       &                &                        & $5.24^{+0.08}_{-0.10}$  & \citet{Mesa2023} & AMES-COND \\
$1030\pm 12$  & $4.00\pm 0.04$ & $-4.75\pm 0.02$      & $4.3^{+2.9}_{-1.3}$     & \citet{DeRosa2023} & AMES-COND, AMES-DUSTY \\
                &                & $-4.81\pm 0.13$      & $3.2^{+0.7}_{-0.6}$     & \citet{Franson:2023arXiv}& empirical BC \\
&&&&&\citet{Filippazzo2015}\\
$789_{-20}^{+22}$ & 3.7 & $-5.22\pm 0.04$ & $2.8^{+0.6}_{-0.5}$ & \citet{Zhang2023} & petitRadTRANS\\
&&&&&\citet{Molliere2019}\\
%800-1100        &                &                        &                           & Palma-Bifani et al. subm.\\
\hline
%$^1$Adopted
\end{tabular}
\end{table*}
\renewcommand{\arraystretch}{1}

Table~\ref{tab:comparison} summarises the parameters for AF Lep b obtained in various papers. We should note that these parameters are not entirely consistent. For instance, the mass and gravity listed by \citet{DeRosa2023} produce a radius of 1.06~R$_{\rm Jupiter}$, which is slightly lower than expected for the mass and age of the planet. \citet{Zhang2023} examined possible inconsistencies in the model atmospheres and their implications. While their analysis for this object is very thorough,  we re-derived some of the relevant quantities in order to be consistent with that for the other sub-stellar companions in the BPMG. In our analysis, luminosities $\log{L/L_\odot}$ are obtained from the $K$-band absolute magnitudes using the bolometric corrections for young ultra-cool dwarfs by \citet{Filippazzo2015}. Radii $R$ are obtained by comparison with the AMES COND evolutionary tracks \citep{Baraffe1998} for an age of 20 Myr. Temperatures are consistent with these values. Errors are obtained by propagating the photometric uncertainties. The effective temperature and luminosity obtained in our analysis of AF Lep are consistent within the errors with those of \citet{Zhang2023}, but we notice that we may slightly overestimate the temperature and luminosity of AF Lep b.

We adopted the dynamical mass obtained by \citet{Franson:2023arXiv} that used a more extensive data set than those considered by \citet{Mesa2023} and \citet{DeRosa2023}. This extensive data set was also considered by \citet{Zhang2023}, who, however, derived a slightly lower mass, albeit within the error bars. This is due to the adoption by \citet{Zhang2023} of a lower mass for the primary star, which in turn is a reflection of the sub-solar metal abundance obtained in their analysis. However, this last analysis may be questionable. In fact, it is well known that the metal abundance of young stars is often underestimated due to the impact of their strong activity on the structure of the atmospheres \citep{Baratella2020}. Analyses that take this into account generally produced solar-like values. We think that these small inconsistencies may be attributed to the uncertainties still existing both in data and in models and are reasonably represented by the error bars adopted in this paper.

For completeness, in Table~\ref{tab:comparison} we also give the value of the gravity that can be obtained combining the luminosity, radii, and masses obtained this way. We notice that these parameters are consistent within the errors with those of \citet{Zhang2023}.

\section{Comparison between dynamical and evolutionary masses}

\begin{figure}[ht]
\centering
\includegraphics[width=\linewidth]{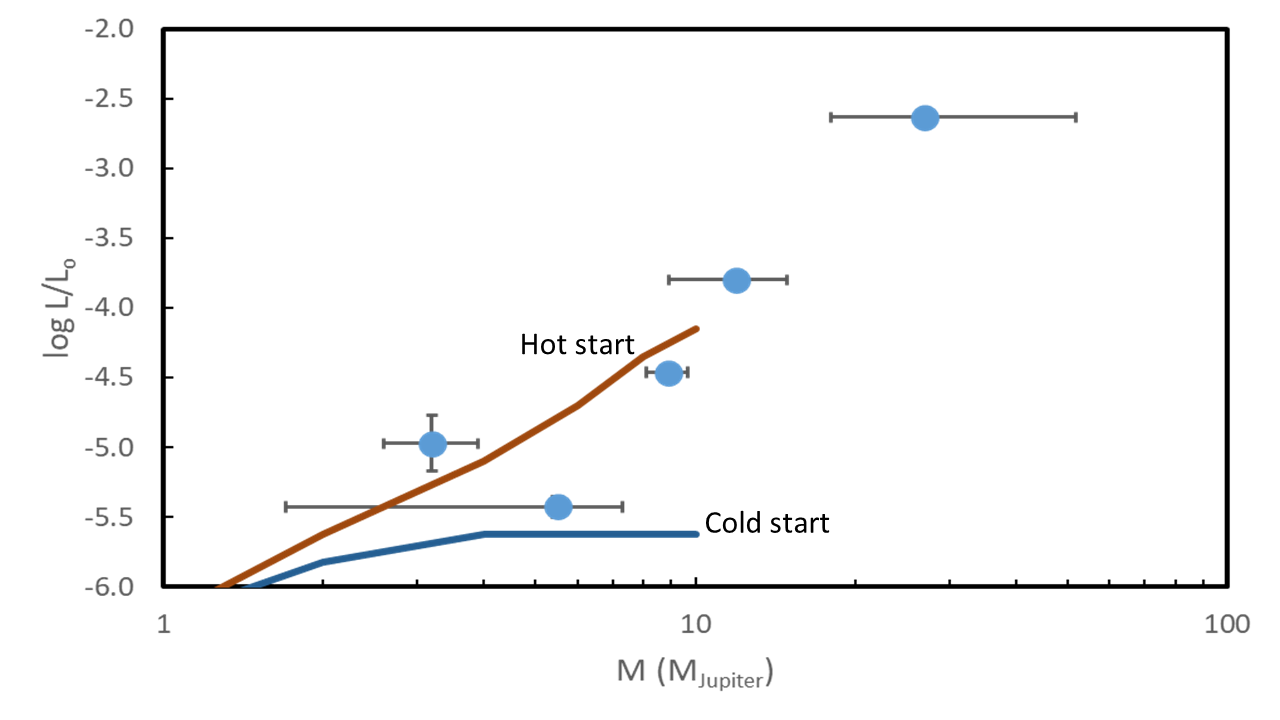}
\caption{Dynamical mass-luminosity relation for sub-stellar objects detected in the BPMG compared with the predictions of hot- (brown line) and cold-start (blue line) models by \citet{Marley2007} for an age of 20 Myr.
}
\label{fig:mass_lum0}
\end{figure}

\begin{figure}[ht]
\centering
\includegraphics[width=\linewidth]{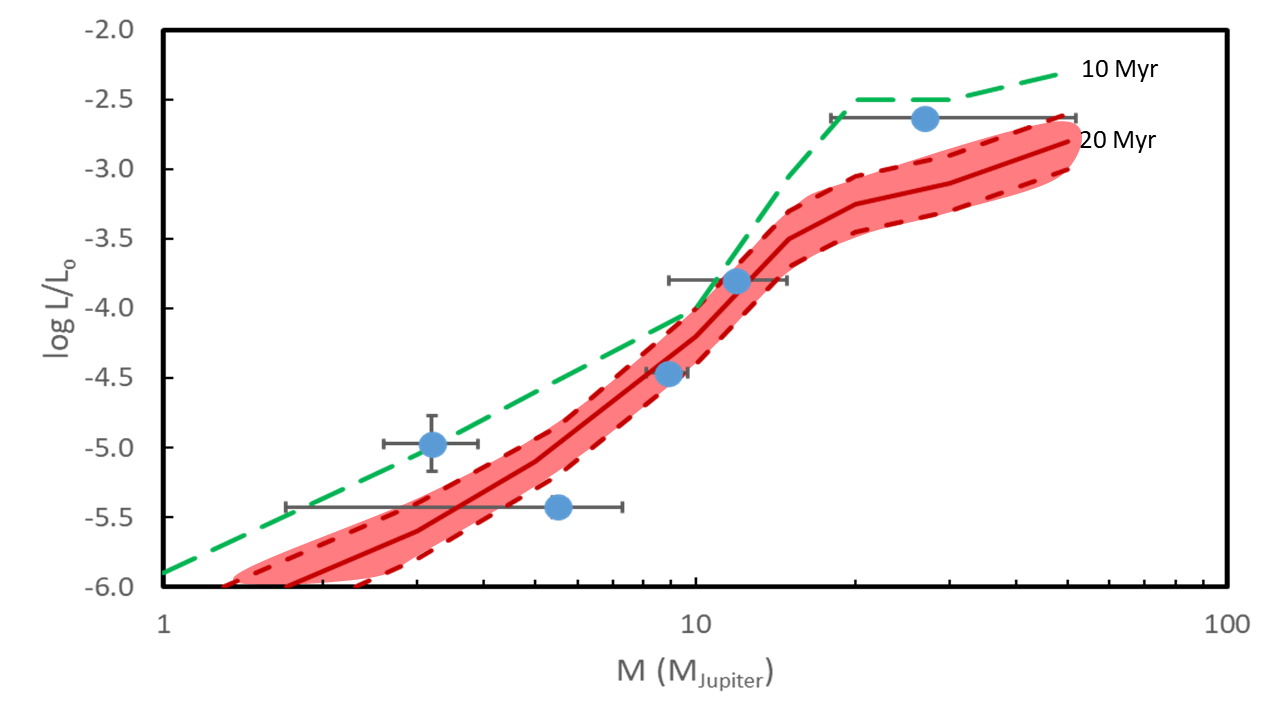}
\caption{Dynamical mass-luminosity relation for sub-stellar objects detected in the BPMG. The solid red and the dashed green lines are averages of the predictions by models of \citet{Mordasini2017} for ages of 20 and 10 Myr, respectively. The shaded red region between the dotted red lines represents the range of values that are expected for a 20 Myr age, depending on the peculiar evolution of individual objects.
}
\label{fig:mass_lum}
\end{figure}

\begin{table*}[ht]
\caption[]{Parameters for sub-stellar companions in the BPMG. Luminosities $\log{L/L_\odot}$ are obtained from the $K$-band absolute magnitudes using the bolometric corrections by \citet{Filippazzo2015}. Errors are obtained by propagating the photometric uncertainties.}
  %Radii $R$ are obtained by comparison with the AMES evolutionary tracks \citep{Allard2001} for an age of 20 Myr. Temperatures are consistent with these values. 
  \label{t:parameter}
  \begin{tabular}{lccccl}
  \hline
%HIP  & Others & Comp & \teff & $R$ & $\log{L/L_\odot}$ & $M$ & Ref \\
%     &   &  & (K)  & (R$_\odot$) &  & (\MJup) & \\
%\hline
%21547 & 51 Eri      & b &  760         & 0.136 & $-5.25\pm 0.05$ & $5.5^{+1.8}_{-3.8}$& \citet{Samland2017, Dupuy2022} \\
%25486 & AF~Lep      & b &  $789_{-20}^{+22}$ & 0.132 &  $-5.22\pm 0.04$ & $3.2^{+0.7}_{-0.6}$& \citet{Franson:2023arXiv, Zhang2023} \\ 
%27321 & $\beta$~Pic & b &  1600        & 0.153 & $-3.86\pm 0.06$ & $11.9\pm 3.0$ &\citet{Nowak2020} \\ 
%      &             & c &  1250        & 0.143 & $-4.35\pm 0.05$ & $8.9\pm 0.8$ &\citet{Nowak2020} \\
%92680 & PZ~Tel      & B &  2700        & 0.22 & $-2.63\pm 0.07$ &$27^{+25}_{-9}$& \citet{Franson2023b} \\ 
%HIP  & Others & Comp & \teff & $R$ & $\log{L/L_\odot}$ & $M$ & Ref \\
%     &   &  & (K)  & (R$_\odot$) &  & (\MJup) & \\
%\hline
%21547 & 51 Eri      & b &  688$\pm$38         & 0.136$\pm$0.003 & $-5.43\pm 0.07$ & $5.5^{+1.8}_{-3.8}$& \citet{Samland2017, Dupuy2022} \\
%25486 & AF~Lep      & b &   908$\pm$123       & 0.132$\pm$0.003 &  $-4.97\pm 0.20$ & $3.2^{+0.7}_{-0.6}$& \citet{Franson:2023arXiv} \\ 
%27321 & $\beta$~Pic & b &  1656$\pm$24       & 0.153$\pm$0.001 & $-3.80\pm 0.02$ & $11.9\pm 3.0$ &\citet{Nowak2020} \\ 
%      &             & c &  1169$\pm$34        & 0.143$\pm$0.001 & $-4.46\pm 0.05$ & $8.9\pm 0.8$ &\citet{Nowak2020} \\
%92680 & PZ~Tel      & B &  2703$\pm$88        & 0.220$\pm$0.004 & $-2.63\pm 0.04$ &$27^{+25}_{-9}$& \citet{Franson2023b} \\ 
HIP  & Others & Comp & $\log{L/L_\odot}$ & $M$ & Ref \\
     &   &  & & (\MJup) & \\
\hline
21547 & 51 Eri      & b & $-5.43\pm 0.08$ & $5.5^{+1.8}_{-3.8}$& \citet{Samland2017, Dupuy2022} \\
25486 & AF~Lep      & b & $-4.97\pm 0.20$ & $3.2^{+0.7}_{-0.6}$& \citet{Franson:2023arXiv} \\ 
27321 & $\beta$~Pic & b & $-3.80\pm 0.05$ & $11.9\pm 3.0$ &\citet{Nowak2020} \\ 
      &             & c & $-4.46\pm 0.07$ & $8.9\pm 0.8$ &\citet{Nowak2020} \\
92680 & PZ~Tel      & B & $-2.63\pm 0.07$ &$27^{+25}_{-9}$& \citet{Franson2023b} \\ 
  \hline
  \end{tabular}
\end{table*}

\begin{figure*}[ht]
\centering
\includegraphics[width=8.8truecm]{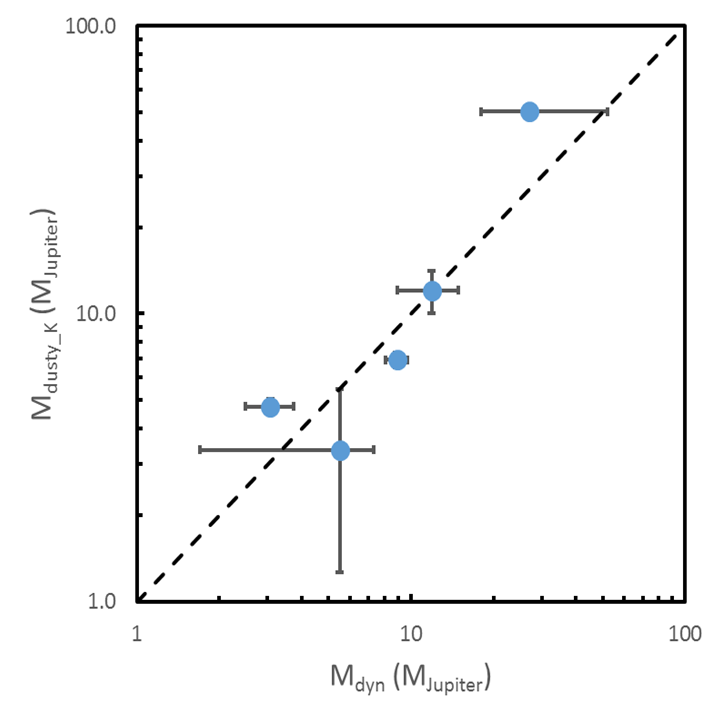}
\includegraphics[width=8.8truecm]{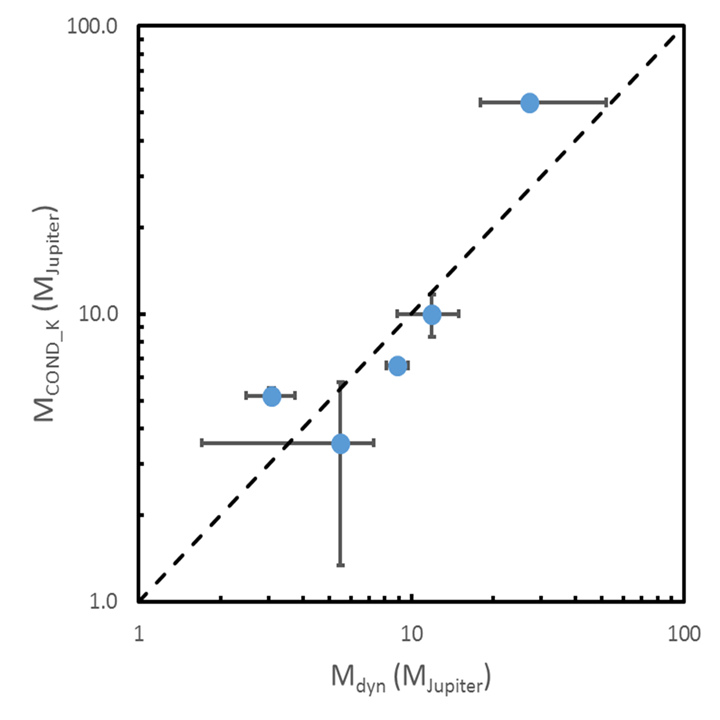}
\includegraphics[width=8.8truecm]{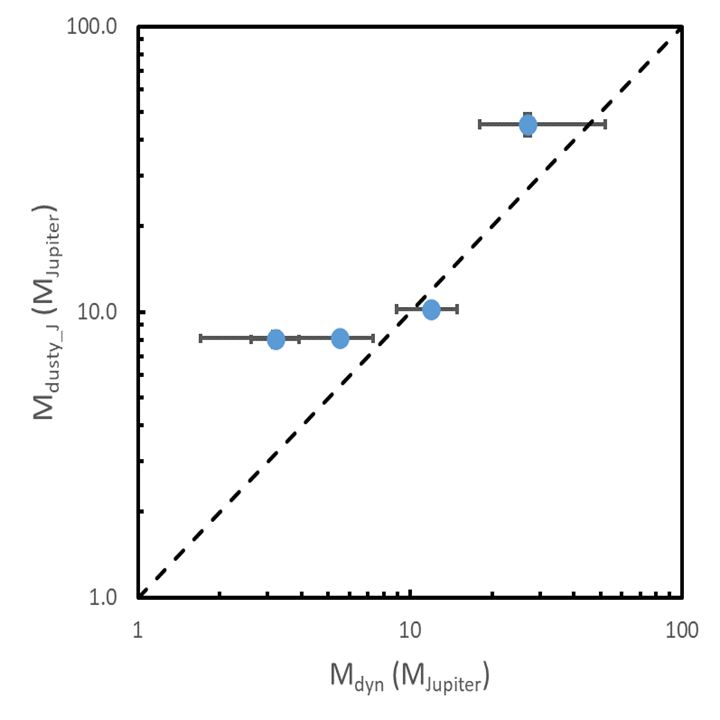}
\includegraphics[width=8.8truecm]{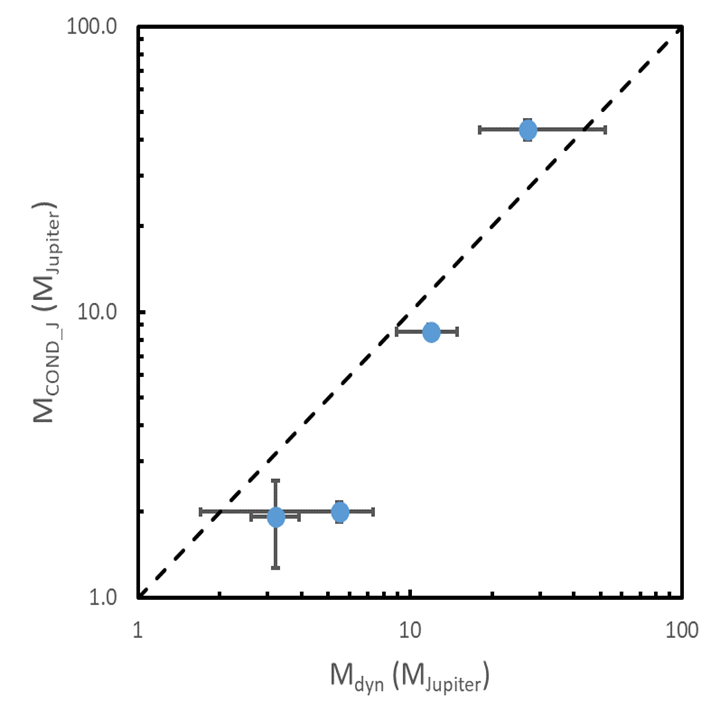}
\caption{Comparison between masses obtained from dynamics and those estimated from evolutionary models for the sub-stellar objects detected in the BPMG using the $K$ magnitude (upper row) and the $J$ magnitude (lower row). The left panels are for masses obtained from photometry using the AMES-DUSTY evolutionary models; the right ones are for masses obtained using the AMES-COND evolutionary models. The dashed lines are for equality.
}
\label{fig:masses}
\end{figure*}

We can use existing data of the dynamical masses of the sub-stellar companions to derive a mass-luminosity relation for the sub-stellar objects in the BPMG. We note here that the age of the BPMG is close to 20 Myr; age estimates \citep{Barrado1999, Mamajek2014, Binks2014, Miret-Roig2020, Couture2023} that used a variety of methods range from 18.5 to 22 Myr. Since they are not directly available, luminosities $\log{L/L_\odot}$ were obtained from the $K$-band absolute magnitudes using the empirical bolometric corrections by \citet{Filippazzo2015}. We give the relevant data in Table~\ref{t:parameter}. We compare the observed mass-luminosity relation with the expectations of hot- and cold-start models by \citet{Marley2007} (see Figure~\ref{fig:mass_lum0}). We remind the reader that we may overestimate the luminosity of AF Lep b, though at the limit of the error bar. This comparison clearly shows that hot-start models match the observational points much better. Figure~\ref{fig:mass_lum} compares the observational data with the predictions of the theoretical core accretion models by \citet{Mordasini2017}. These models actually predict a range of possible values, depending on the exact history of every single planet; so rather than a single mass-luminosity relation, an intrinsic scatter of the luminosities is expected at each mass and age. In the figure, this is represented by the shaded area between the dashed lines. The comparison between models and observations is reasonably good, in view of the large uncertainties in individual points.

The models by \citet{Baraffe1998} that use the AMES line list are popularly used to derive evolutionary masses for targets lacking a dynamical mass. We then show in Figure~\ref{fig:masses} the comparison between dynamical masses and those that are derived from the application of these evolutionary models. We show data obtained both with cloudy (AMES-DUSTY: \citealt{Chabrier2000}) and clear (AMES-COND: \citealt{Allard2001}) model atmospheres, when using the $K$ magnitude. With the addition of AF Lep b, this is now possible for a total of five companions in the BPMG, the others being the two planets of $\beta$~Pic, that of 51~Eri, and the brown dwarf (BD) PZ~Tel~B. For the time being, both models (that correspond to 'hot-start' models) pass this test. However, both sets of isochrones perform poorly when using the $J$ magnitudes. For this band, the DUSTY models overestimate the masses while the COND ones underestimate them. We empirically found that an average of the two values reproduces the dynamical masses well. This result implies that the models give a poor representation of the $J-K$ colour of the planets. We discuss this point in Section 4.

%\subsection{Relation between masses and spectral types}

%\begin{figure}[ht]
%\centering
%\includegraphics[width=\linewidth]{spectrum.png}
%\caption{Relation between mass and spectral type for sub-stellar objects in the BPMG
%}
%\label{fig:Spectra}
%\end{figure}

%The relation between masses and spectral types of these five companions provides a test on model atmospheres. This is shown in Figure~\ref{fig:Spectra}. We notice that the spectral type is closely related to the temperature of the companion and that AF Lep b fits quite well in the relation established by the other known sub-stellar objects in the BPMG, being intermediate both in mass and spectral type between 51~Eri~b and $\beta$~Pic~b. 

%\FloatBarrier

\section{Colour-magnitude diagram for young planets and free-floating objects}

\subsection{Observational data}

\begin{figure}[ht]
\centering
\includegraphics[width=\linewidth]{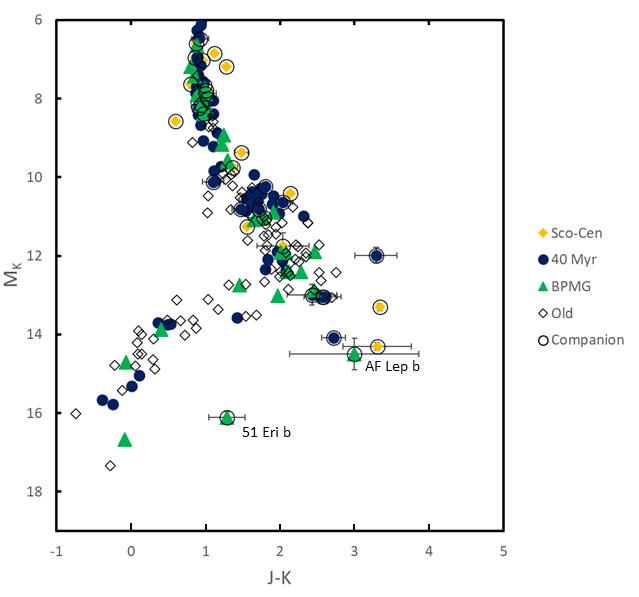}
\caption{(M$_K$-$J-K$) colour-magnitude diagram for sub-stellar objects in BPMG (green-filled triangles). Sub-stellar object members of Sco-Cen (orange diamonds), young nearby associations with ages in the range of 40-50 Myr (blue-filled circles), and older ones (open blue diamonds) are also plotted. Blue circles mark objects that are companions of more massive objects.} %Blue and red solid lines are the predictions of 20-year-old AMES-COND and AMES-DUSTY isochrones.}
\label{fig:cmd}
\end{figure}

\begin{figure}[ht]
\centering
\includegraphics[width=\linewidth]{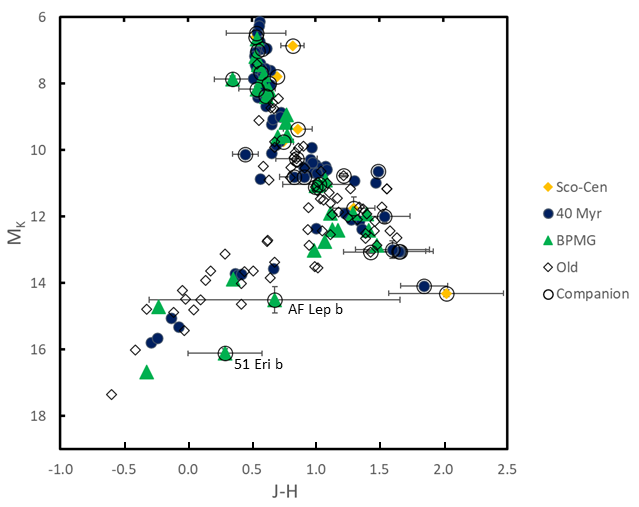}
\caption{(M$_K$,$J-H$) colour-magnitude diagram for sub-stellar objects in the BPMG (green filled triangles). Sub-stellar objects members of the Sco-Cen (orange diamonds), of young nearby associations with ages in the range of 40-50 Myr (blue-filled circles) and of older ones (empty blue diamonds) are also plotted. Blue circles mark objects that are companions of more massive objects. }
\label{fig:cmd-jh}
\end{figure}

\begin{figure}[ht]
\centering
\includegraphics[width=\linewidth]{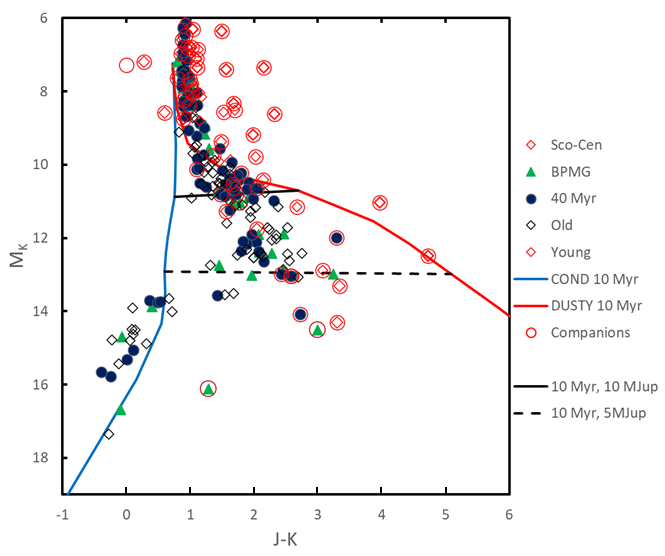}
\caption{Same as Figure~\ref{fig:cmd}, but with the inclusion of associations younger than 10 Myr (red circles). Red circles mark objects that are companions of more massive objects. Solid blue and red lines are the predictions of AMES-COND and AMES-DUSTY isochrones with an age of 10 Myr. The solid and dashed black lines connect the points corresponding to the COND and DUSTY AMES isochrones for an age of 10 Myr and masses of 10 and 5 \MJup, respectively. {These nearly horizontal iso-mass lines shown here are representative of other masses and ages that span the relevant range of the analysis.}%The violet dashed line connects 5~M$_/{\rm Jupiter}$ along these isochrones  and corresponds to an effective temperature of 1193~K
}
\label{fig:cmd_all}
\end{figure}

\begin{table*}
\caption{Ages for associations and moving groups (in Myr)}
\begin{tabular}{lcccccc}
\hline
Reference & BPMG & Columba &	Carina	& Tuc-Hor	& Argus	& AB Doradus\\
\hline
\multicolumn{6}{c}{Isochrones}\\
\hline
\citet{2015MNRAS.454..593B} & 24 & 42 & 45 & 45 & 58 & 149 \\
\citet{Booth2021}           &    &    & 13 &    &    &     \\
\hline
\multicolumn{6}{c}{Lithium}\\
\hline
\citet{Barrado2004}	        &    &    &    &    & 50 &     \\	
\citet{Mentuch2008}         & 21 &    &    & 27 &    &$>$45 \\
\citet{Binks2014}           & 21 &    &    &    &    &      \\
\citet{Malo2014}            & 26 &    &    &    &    &      \\
\citet{Shkolnik2017}        & 22 &    &    &    &    &      \\
\citet{Schneider2019}       & 22 &	  & 22 &    &    &      \\
\citet{Wood2023}            &    &    &	40 &    &    &      \\			
\hline
\multicolumn{6}{c}{Kinematics}\\
\hline
\citet{McCarthy2014}        &    &    &    &    &    & 125 \\
\citet{Miret-Roig2020}      & 13 &$>$40&$>$28&$>$28&37&    \\	
\citet{Miret-Roig2020}      & 19 &    &    &    &    &     \\
\citet{Booth2021}           &    &    & 20 &    &    &     \\
\citet{Kerr2022}            &    & 26 & 26 & 46 &    &     \\
\citet{Couture2023}         & 20 &    &    &    &    &     \\
\hline								
Mean                        & 21 & 36 & 28 & 37 & 48 & 137 \\
Standard deviation          &  4 &  8 & 11 & 11 & 10 &  17 \\
\hline
\end{tabular}
\label{t:age1}
\end{table*}

\begin{table}
\caption{Ages for additional associations and moving groups (in Myr)}
\begin{tabular}{lcl}
\hline
Associations &	Age	 & Reference \\
\hline
Taurus	       & 1-2 & \\	
Chamaleon      & 1-2 & \\	
$\epsilon$ Cha & 11  & \citet{2015MNRAS.454..593B} \\
TWA            & 10  & \citet{2015MNRAS.454..593B} \\
Upper Scorpius  & 8-12 &	\citet{Pecaut2016}  \\
Upper Centaurus-Lupus	 & 12-16 &	\citet{Pecaut2016}  \\
Lower Centaurus-Crux	 & 16 & \citet{Pecaut2016}  \\
Carina Near & 200 & \citet{Zuckerman2006} \\
%UMA & 414 & \citet{Jones2015} \\
%Hyades & 680 & \citet{Gossage2018} \\	
\hline
\end{tabular}
\label{t:age2}
\end{table}

The location of sub-stellar objects in the BPMG in the colour-magnitude diagram provides further insights into their properties. To give more statistical weight to our results, we considered both sub-stellar companions and free-floating objects belonging to a number of young populations, in addition to the BPMG: Sco-Cen, Columba, Carina, Argus, Tucana-Horologium, Taurus, Chamaleon, TWA, AB Doradus, and Carina Near. For each of the objects, we checked membership to the respective groups considering the parallaxes, proper motion, and, when available, radial velocities, and using the online BANYAN $\Sigma$ code \citep{Gagne2018}\footnote{\url{https://www.exoplanetes.umontreal.ca/banyan/banyansigma.php}}. References for the data of the individual objects can be found in Appendix~\ref{sec:photometry}. In addition, we considered data for sub-stellar objects in the Upper Scorpius association from \citet{Lodieu2006} and \citet{Bouy2022}. 

Interpretation of this photometry requires an estimate of the age of the individual objects. In Tables~\ref{t:age1} and~\ref{t:age2}, we report a number of literature determination of ages for the various associations and moving groups considered in this paper. The values we adopted are the straight averages. In this work, we did not consider the ages by \citet{Ujjwal2020} because they are much lower than and uncorrelated with other estimates. In addition, \citet{Gagne2018wd} reported an age of $117\pm 26$ Myr for the AB Dor using the massive white dwarf GD 50. While we did not use this estimate here, it is consistent within the errors with the value we adopted.

We show the $(M_K,J-K)$ and the $(M_K,H-K)$ colour magnitude diagrams for these objects in Figure~\ref{fig:cmd} and Figure~\ref{fig:cmd-jh}, respectively. These figures show that the most massive sub-stellar companions of the BPMG (those with $M_K<12.5$) are indistinguishable in this diagram from the free-floating objects and from members of the other associations of different ages if they are older than 10-15 Myr. This agrees with earlier findings that the colour-magnitude diagram of sub-stellar objects does not change much for ages less than a few hundred million years \citep{Faherty2016}. However, the case is different for fainter objects. While the L-T transition\footnote{We do not actually use spectral types throughout this paper; we use the term L-T transition because it is the appearance of very strong molecular bands - faint in L-type spectra and prominent in T-type ones which causes the change in $J-K$ colour from very red to blue.} occurs at magnitudes in the range $M_K\sim 13$ for free-floating objects (irrespective of their age, at least in this range), fainter companions are still on the L-sequence down to $M_K\sim 14.5$. There are in fact five planets with $13.3<M_K<15$ and $J-K>2.5$; they are AF Lep b, HR8799b, HD95086b, TYC 8998-760-1c, PDS-70c. Admittedly, PDS-70c may be reddened by the circumstellar  (projected towards the planet) and/or circumplanetary disc, so we preferred not to consider it. Still, there are at least four extremely red planets that have no counterpart among the free-floating objects. On the other hand, there is no known planetary companion with $J-K<1.5$ and $13.3<M_K<15$, which is a region populated by more than 20 young free-floating objects. Again, this is not very sensitive to age; in fact, the same occurs for companions in the other associations considered in the plot.

We must caution that this might in part be due to some selection effect. Indeed, detection of such faint and red objects is very difficult and they may have been missed in the surveys looking for low-mass free-floating objects in these young associations. For instance, very recently \citet{Schneider2023} announced the discovery of an extremely red free-floating object in the BPMG (CWISE J050626.96+073842.4) with $M_K=12.99$ and $J-K=2.97$. In the (M$_K$,$J-K$) colour-magnitude diagram, this object occupies a position very similar to those of the inner planets of HR8799, but it still is brighter than the four planets considered above. In addition, we notice that free-floating objects with a mass of 5-6~\MJup in the BPMG such as 2MASS J08195820-0335266 and CFBDS J232304-015232\footnote{The masses for these two objects are 6.16$\pm$ 0.55~\MJup and 5.13$\pm$ 0.44~\MJup, using the approach described later in this section.} have a late-T spectral type and a bright $M_K$\ magnitude of around 14 \citep{Zhang2021}. These objects are roughly as massive as 51 Eri b and AF Lep b, which are a factor of ten fainter in the $J$ band. The comparison with slightly older objects in other associations suggests that these two free-floating T-objects are not exceptional. \citet{Liu2016} already noticed the existence of systematic differences between the colour-magnitude diagram of young free-floating and companion sub-stellar objects.

\subsection{Remapping data in a temperature - relevance of dust plane}

\begin{figure}[ht]
\centering
\includegraphics[width=\linewidth]{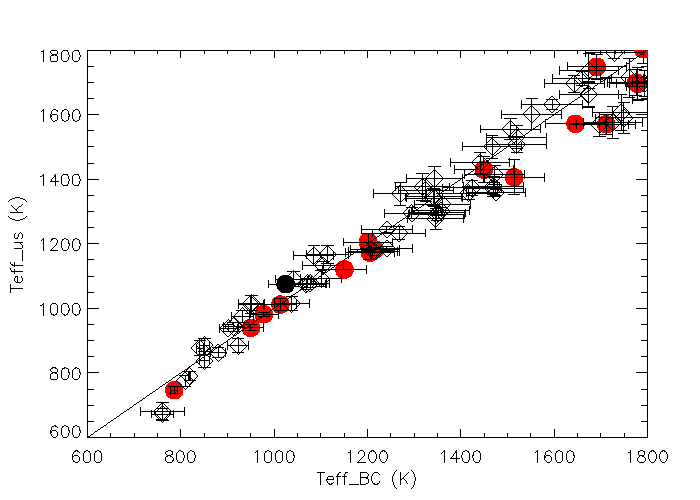}
\caption{Comparison between \teff obtained from luminosities derived from $K-$ band magnitudes and bolometric corrections from \citet{Filippazzo2015}, combined with evolutionary radii, with the \teff obtained from our remapping approach. Filled red circles represent companions within 1000 au, filled black circles represent companions outside 1000 au, and open diamonds represent free-floating objects or very wide companions (separation $>1000$~au). The solid line represents equality.}
\label{fig:teff-comp}
\end{figure}

\begin{figure}[ht]
\centering
\includegraphics[width=\linewidth]{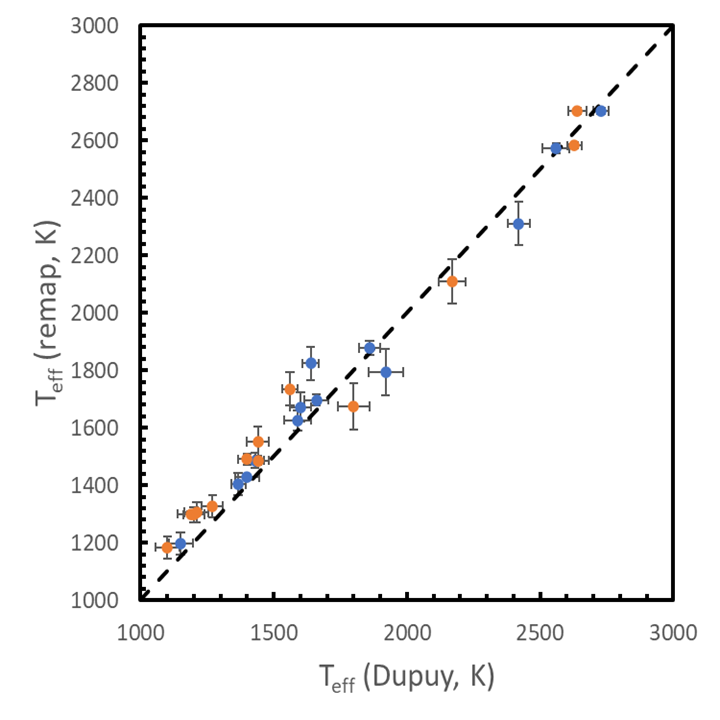}
\caption{Comparison between \teff values obtained by \citet{Dupuy2017} with those obtained from our remapping approach for their sample of BD binaries. Blue symbols are primaries, orange ones are secondaries. }
\label{fig:teff-dupuy}
\end{figure}

\begin{figure}[ht]
\centering
\includegraphics[width=\linewidth]{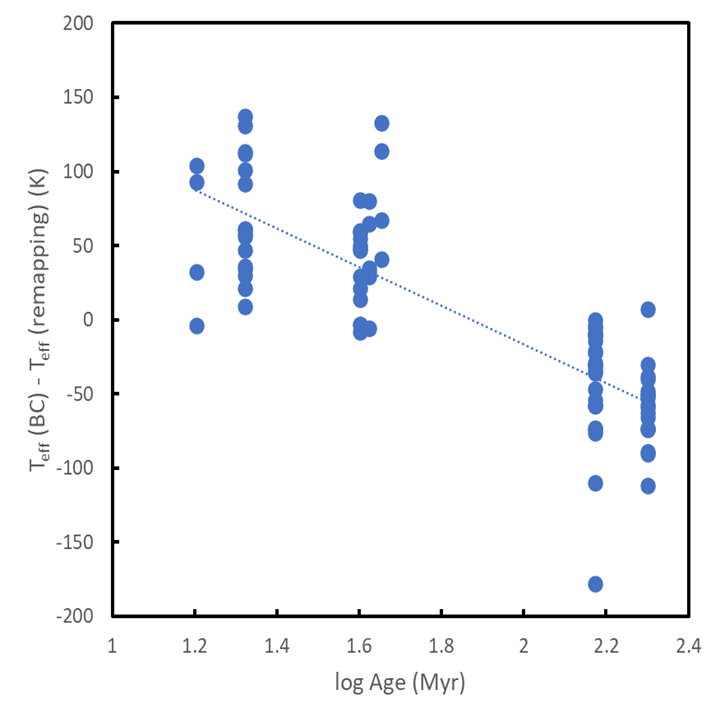}
\caption{Run of offset between \teff from BC and remapping as a function of the age for sub-stellar objects with \teff$<1800$ K. The dashed line corresponds to Eq. (1) in the text. }
\label{fig:log_age_dteff}
\end{figure}

%\begin{figure}[ht]
%\centering
%\includegraphics[width=\linewidth]{Teff_rJ-H.eps}
%\includegraphics[width=\linewidth]{Teff_rJ-K.eps}
%\includegraphics[width=\linewidth]{Teff_r.eps}
%\caption{Remapping of position of stars from the colour-magnitude diagram (cmd) into the plan \teff vs dust relevance parameter $r$ (see text) for sub-stellar objects with ages in the range $10-200$~Myr. The upper panel show results obtained using the $(M_J, (J-H))$ cmd, the intermediate panel using the $(M_J, (J-H))$ cmd, and the lower panel the average results from these diagrams. In all panels, filled circles are companions and open diamonds are free floating objects or very wide companions (separation $>1000$~au). The solid and dashed lines represent the expectations for AMES-COND and AMES-DUSTY models, respectively.}
%\label{fig:teff-r}
%\end{figure}

\begin{figure}[ht]
\centering
\includegraphics[width=\linewidth]{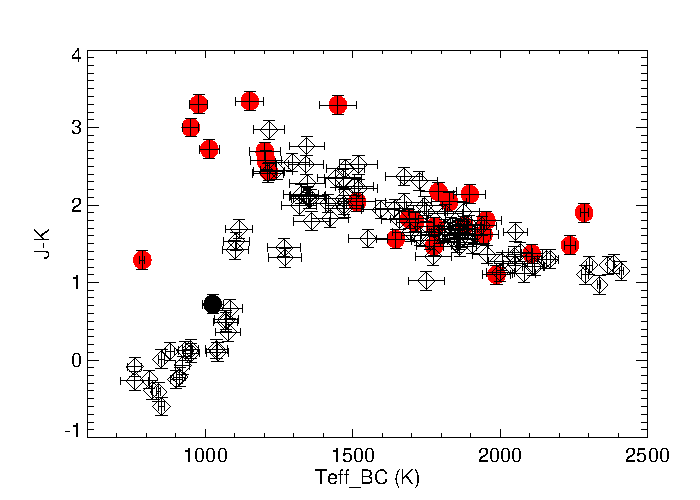}
\caption{Run of $J-K$ colour as a function of \teff obtained from $K$ magnitude, bolometric corrections by \citet{Filippazzo2015} and radii from evolutionary models. Filled red circles are companions within 1000 au, filled black circles show companions outside 1000 au, and open diamonds show free floating objects or very wide companions (separation $>1000$~au).}
\label{fig:Teffbc_J-K}
\end{figure}

\begin{figure}[ht]
\centering
\includegraphics[width=8.8truecm]{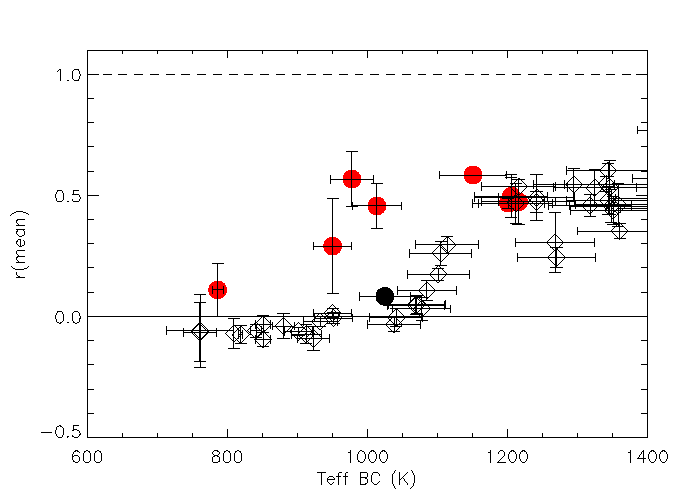}
\includegraphics[width=8.8truecm]{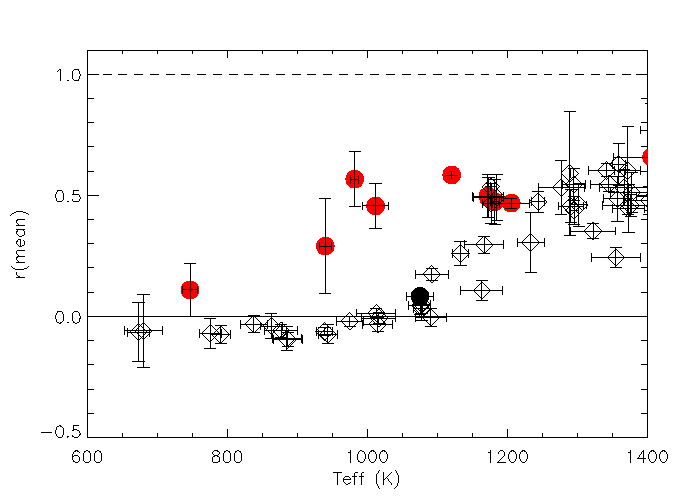}
\caption{Remapping of positions of stars from colour-magnitude diagram (cmd) into plan \teff versus dust relevance parameter $r$ (see text) for sub-stellar objects with ages in the range $10-200$~Myr. The upper panel shows results obtained using \teff obtained from luminosities derived from the $K$ band magnitudes and bolometric corrections from \citet{Filippazzo2015}, combined with evolutionary radii. The lower panel shows \teff from our remapping approach. In both panels, filled red circles are companions within 1000 au, filled black circles show companions outside 1000 au, and open diamonds show free-floating objects or very wide companions (separation $>1000$~au). The solid and dashed lines represent the expectations for AMES-COND and AMES-DUSTY models, respectively.}
\label{fig:teff-r}
\end{figure}

\begin{figure}[ht]
\centering
\includegraphics[width=\linewidth]{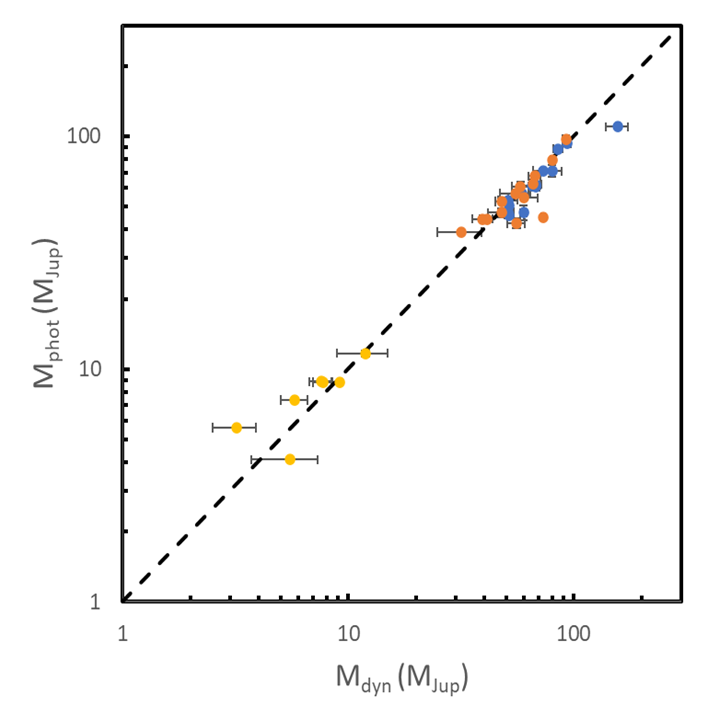}
\caption{Comparison between masses obtained from dynamics and those estimated from evolutionary models for sub-stellar objects. Evolutionary masses were obtained from the AMES isochrones using the approach described in this paper. Yellow symbols represent planetary companions; blue-filled circles show primaries and orange symbols show secondaries from the binary BD sample of \citet{Dupuy2017}. The dashed line is for equality.
}
\label{fig:masses-2}
\end{figure}

We remind the reader that while alternative scenarios have been proposed (see \citealt{Tremblin2016}), the L-T transition is attributed by most authors to the settling of clouds in the atmospheres that are thought to be very abundant in L-type objects. This is shown by a comparison of the observed location of stars in the (M$_K$,$J-K$) colour-magnitude diagram with AMES-COND and DUSTY isochrones (see Figure \ref{fig:cmd_all}). As mentioned previously, these isochrones use the same evolutionary tracks (that assume a hot start) but different model atmospheres. By definition, the COND atmospheres have no clouds. On the other hand, DUSTY atmospheres are very rich in dust because they assume no settling - while some dust sedimentation is generally expected \citep{Marley2010, Morley2012}. While the L-sequence is close to the AMES DUSTY isochrone, the late T sequence is close to the AMES-COND one. For free-floating objects, the L-T transition occurs at an absolute $K$ magnitude between 13 and 14, which, for the age of the BPMG corresponds to a mass of about 5~\MJup. An extension of the survey to older associations (AB Doradus and Carina-Near) shows that the magnitude at which the L-T transition occurs is rather stable over a quite wide range of ages (\citealt{Liu2016}; see also Figure~\ref{fig:cmd_all}). % and \ref{fig:cmd-jh_all}).  

To show the relation between the effective temperature and the transition from cloudy to clean atmospheres, we remapped the colour-magnitude diagram (cmd) into an effective temperature - relevance of dust plane. In our approach, this last effect is represented by a parameter $r$. To obtain this parameter, we started from the AMES-COND and AMES-DUSTY isochrones. As mentioned earlier, these isochrones correspond to the same internal model (hence, the same age, mass, temperature, and luminosity), but they use different model atmospheres that project into very different sequences in the near infrared (NIR) cmds. For a given age, for any value of \teff (or mass) we may then define two points in the cmd corresponding to the COND and DUSTY isochrones. By linear interpolation between these points, we can define a new isochrone that corresponds to any arbitrary value of $r$, where $r=0$ for the AMES-COND isochrones and $r=1$ for the AMES-DUSTY one. Hence, a higher value of $r$ qualitatively corresponds to higher dust relevance in the atmosphere. Since the AMES-COND and AMES-DUSTY models are not perfect, very dusty or clean atmospheres do not actually correspond to $r=1$ or 0.  In particular, while the AMES-COND models are quite good representations of a clean atmosphere in this context, the dust-rich atmospheres are much less red than expected from AMES-DUSTY models and correspond to a value of $r\sim 0.5$ rather than $r\sim 1$. However, the relative scale is still valid, at least for \teff$<1800$~K. At high temperatures, colour predictions with AMES-COND and AMES-DUSTY models are quite similar, and the value of $r$ becomes uncertain. This result has a weak dependence on age that acts on the radius and then on the magnitude, and it should thus be taken into account. Through this remapping, the location in the NIR cmd for each object (its absolute magnitude and colour) corresponds to a pair of values of \teff and $r$. 

In practice,  we ran a Monte Carlo procedure for each star extracting 100 random values of age from a Gaussian distribution with the appropriate mean values and standard deviations. For each of these random sets of ages, we also extracted values of colour and magnitude again with Gaussian distributions with means equal to the best value and standard deviations equal to the error appropriate to the observation of star. For each run of the Monte Carlo procedure, we then constructed maps of colour and magnitude as a function of mass and value of $r$ with a very fine grid in both quantities. We then found where the quadratic sum of the differences between the predicted colour and absolute magnitude and the 'observed' values given by the Monte Carlo procedure described above are minimised. The final best values of mass and $r$ are the mean of the values obtained this way, and the error is the standard deviation of these values. When constructing the maps, we considered values of $r$ in the range of $-0.8 \div 1.9$; values outside the range $0 \div 1$ were obtained by simple linear extrapolation of those corresponding to COND and DUSTY isochrones.

We considered this procedure for both the ($M_K,(J-H)$) and ($M_K,(J-K)$) cmds and we called the two values of $r$ obtained for each diagram $r(J-H)$ and $r(J-K)$. We also obtained a value that we called $r(mean)$, which is the weighted average of the two values.

Figure~\ref{fig:teff-comp} compares the effective temperatures obtained using this approach with \teff obtained from luminosities derived from the $K-$ band magnitudes and bolometric corrections from \citet{Filippazzo2015}, combined with evolutionary radii for young objects. In addition, we repeated the same analysis for the objects listed by \citet{Dupuy2017} (see Figure~\ref{fig:teff-dupuy}). The brown dwarf binaries considered in this last paper are mainly old ones (age $>271$ Myr). Moreover, all \teff values by \citet{Dupuy2017} are $>1000$ K, and among those with ages $<1$ Gyr, the coolest one is Gl 417C with \teff=1560 K. In both cases, there is an excellent agreement for the stars cooler than 1800 K, while our procedure gives large errors for higher \teff values. This is because at \teff$>1800$~K DUSTY and COND isochrones are essentially coincident, and the value of the dustiness parameter $r$ has large errors. 

We derived internal errors on the \teff from the remapping procedure considering photometric errors and comparing results from $J-H$ and $J-K$. We obtained a mean quadratic value of $\pm 34$ K for stars with \teff$<1800$ K, while errors are as high as $\pm 175$ K for stars warmer than this limit. Since both these results depend on the $J$ magnitude, they are not independent of each other. Hence, tests with temperatures obtained using different methods, e.g. from the $K$-magnitudes, bolometric corrections, and radii from models, are more meaningful. We compared the \teff obtained from remapping with those obtained from bolometric magnitudes. If we limit ourselves to objects with \teff$<1800$ K, temperatures obtained by remapping are, on average, lower by $9\pm 7$ K, with an r.m.s. of the difference of 65 K. This corresponds fairly well to the internal errors of 47 and 34 K obtained for the temperatures from bolometric magnitudes and remapping, respectively. We notice that for stars cooler than 1800 K, the residuals are a clear function of age (see Figure \ref{fig:log_age_dteff}), and they are represented by the following relation\footnote{The age distribution of the sub-stellar objects considered in this paper actually consist of two groups: young objects with ages $<50$ Myr and older ones with ages $>140$ Myr. This fact is responsible for the apparent presence of two separate sequences in Figure~\ref{fig:teff-comp}.}:
\begin{equation}
T_{\rm eff\, remapping}-T_{\rm eff\, BC} = 130.35 \log{(Age/Myr)} - 244\, {\rm K}
\end{equation}
Root mean square residuals around this relation are only $\pm 37$ K, which is consistent with the internal errors of each of the two relations. This fact can be attributed to our use of a unique BC relation (the one appropriate to young BDs) from \citet{Filippazzo2015}, while the same authors acknowledge that this should depend on age.  This suggests that in this temperature range, the two methods provide nearly equally accurate \teff values.

We also compared the \teff obtained by our approach with those obtained by \citet{Dupuy2017}, which also used $K$ magnitudes, bolometric corrections, and radii from models. Most of the stars they considered are warmer than 1800 K. On average our \teff are higher by $37\pm 17$ K; residuals have an r.m.s. of 86 K. Considering that the internal error of \citet{Dupuy2017} is 40 K, we may estimate that our \teff for these stars have errors of 75 K. 

Figure \ref{fig:Teffbc_J-K} shows the run of the $J-K$ colour as a function of the \teff obtained from the $K$ magnitude, the bolometric corrections by \citet{Filippazzo2015} and radii from evolutionary models. This figure (that is actually very similar to the colour-magnitude diagram shown in Figure \ref{fig:cmd-jh} save for the inversion of the axes) shows the expected difference between planets and free-floating objects, with the group of cool and very red planets at \teff$<1100$ K and $J-K\sim 3$, a region where there is no free-floating object. For better insight into the nature of this difference, Figure~\ref{fig:teff-r} shows the results of our remapping into the \teff-$r$ plane for the sub-stellar objects with ages in the range of 10-200~Myr listed in the appendix. Different symbols are used for companions and free-floating objects. As can be seen, as soon as the temperature of a free-floating sub-stellar object falls below $\sim 1200$~K (this is indeed the direction of the evolution of these objects) the $r$ parameter drops, indicating that clouds settle and the atmosphere becomes quite transparent. However, companions (at least those with separation $<1000$~au) behave differently, and cloud settling occurs at a much lower temperature ($<1000$~K). This results in the existence of extremely red objects with $M_J>16.5$ and $J-K>2.7$, such as HR8799b, HD~95086b, and TYC~8998-760-1c, that have no counterparts among free-floating objects. In addition, the atmosphere of AF~Lep~b, with \teff$=789_{-20}^{+22}$ K, still looks very dust-rich (a fact already noticed by \citealt{Zhang2023}). 51~Eri~b is also redder than free-floating objects with the same temperature. 

A secondary but interesting output of the remapping of the sub-stellar objects in the \teff$-r$ plane is the possibility of obtaining a homogeneous set of evolutionary masses for them. This is given in Table~\ref{t:teff-r-mass}. These masses can be compared with the dynamical masses for the planets in the BPMG as well as for those around HR~8799 in the Columba association (\citealt{Zurlo2022}: see Figure~\ref{fig:masses-2}). The agreement is good. We also compared the masses with those of \citet{Dupuy2017} and again found good agreement. On average our masses are lower by $-2.6\pm 1.4$ \MJup, which is less than 5\%; residuals have an r.m.s. of 7.3 \MJup, which agrees fairly well with the combinations of the internal errors of \citet{Dupuy2017} (6.5 \MJup) and from our formulas (1.8 \MJup). However, we notice that our approach requires independent knowledge of the ages. Unluckily, this is not the case for the binaries considered by \citet{Dupuy2017}. The age they gave for a number of their targets is actually derived from fitting models to their observational data (magnitudes and dynamical masses). It is then not surprising that we found extremely good agreement between the masses that we may obtain from photometry and their dynamical masses. For these stars, the result only shows that the two analyses are consistent with each other, but not that the photometric masses are correct.

\subsection{Why the L-T transition for companions occurs at a different temperature from that of free-floating objects}

While the statistics is still limited, the results of this section suggest that temperature and gravity (and then age) are not the only parameters controlling cloud settling. Given the complexity of cloud physics, there are various possible explanations. For instance, the models by \citet{Charnay2018} show that the difference between companions (such as AF Lep b) and free-floating objects of similar mass and age (such as 2MASS J08195820-0335266 and CFBDS J232304-015232) might be obtained if the size of dust grains in the atmospheres of companions is smaller than that in the atmospheres of free-floating objects. Such a difference might perhaps be related to a systematic difference in the chemical composition. Early analysis concluded that young planets that accrete gas from the disc will most likely have a strongly oxygen-depleted atmosphere \citep{Helling2014}. The grain seed formation rate decreases with decreasing oxygen abundance and increasing carbon abundance. This results in fewer cloud particles being formed; grains should rain on denser layers and grow to larger sizes than in O-rich atmospheres. However, the inclusion of pebble \citep{Schneider2021a, Schneider2021b} and collisional \citep{Ogihara2021} accretion substantially revised this conclusion, showing that the atmospheres of giant planets might be highly enriched in volatile elements (CNO). The very high value of the metallicity obtained for AF Lep b by \citet{Zhang2023} and the moderate one for 51 Eri b by \citet{Samland2017} indeed support this.

Alternatively, we may think that rotation is systematically different in companions and free-floating objects. A higher rotation reduces the efficiency of turbulence and the net speed of vertical motions \citep{Brummell1996}; this in turn implies a higher value for the ratio $f_{\rm sed}$\ of the particle sedimentation velocity to the characteristic vertical mixing velocity \citep{Ackerman2001}. The effect is complex because a higher $f_{\rm sed}$ also implies a higher sedimentation radius, but in general, we expect that a higher value of $f_{\rm sed}$\ should correspond to cleaner atmospheres (see also Figure 15 in \citealt{Charnay2018}). Hence, a faster rotation should produce cleaner atmospheres. There is no evidence that companions rotate slower than free-floating objects. For instance, the rotational period estimated for $\beta$~Pic b (8.3 hr; \citealt{Snellen2014}) is actually shorter than the median value of about 1~d found for young (free-floating) brown dwarfs by \citet{Scholz2016}, though it is at the lower end of the range of observed values.

We also notice that the origin of free-floating planets is not yet well established (see discussion in \citealt{2023Ap&SS.368...17M}). Several studies indicate that the observed fraction of these objects outnumbers the prediction of cloud turbulent fragmentation (see e.g. \citealt{2022NatAs...6...89M}) and suggest that many were formed in discs around protostars that were later ejected. The colour of free-floating planets may suggest a preference for their formation through gravitational instability. This might indicate that turbulent fragmentation of discs plays a fundamental role in the genesis of free-floating planets, although other channels of formation are also very likely to occur. If this were true, the different dust-settling temperatures of planets might be related to their formation scenario (core accretion vs. disc instability).   

\FloatBarrier

\section{Conclusions}

AF Lep b is the fourth planet discovered through high-contrast imaging in the $\beta$~Pic moving group, and one of the few extrasolar planets for which dynamical mass and luminosity are available. Consideration of data for this planet strengthens the early conclusion that young massive planets evolve much closer to hot-start models rather than to cold-start ones. The mass-luminosity relation found using the planets in the $\beta$~Pic moving group is in agreement with the most recent formation and evolution models for giant planets in the core accretion scenario. 

Meanwhile, the extensive photometric data gathered recently for sub-stellar objects in young associations and moving groups enables a comparison of the L-T transition occurrences between companions and free-floating objects. These data indicate that the L-T transition occurs at nearly the same magnitude for free-floating objects over quite a large age range (at least up to that of the Hyades) as previously noticed by \citet{Liu2016}. The L-T transition is possibly due to the settling of dust in the atmospheres of sub-stellar objects. Since all objects in the mass range between Jupiter and the hydrogen-burning limit share a similar radius with a small range, this means that the settling of dust occurs at a nearly constant temperature of about 1200 K in free-floating objects, rather irrespective of their mass\footnote{In principle correct the bolometric correction should also be considered here. However, as shown by Figure \ref{fig:teff-r}, the L-T transition of the free-floating objects occurs in a rather narrow range of temperatures between 1100 and 1200 K. The range is even more restricted when we use temperatures obtained considering appropriate bolometric corrections.}. In contrast, the $\beta$~Pic moving group planets that are intermediate between the red (dusty) and blue (clean) sequences such as 51 Eri b and AF Lep b are about two magnitudes fainter in the $J$ band than free-floating objects of presumably the same mass belonging to the same association. This suggests that the L-T transition - and hence the dust settling - occurs at a lower temperature (about 800-1000 K) in sub-stellar companions than in free-floating objects. This feature is probably not unique to this association. Notably, the sequence of sub-stellar companions features very red (that is cool, dust-rich) objects, which are not observed in free-floating objects.

The reason for this difference between free-floating and companion sub-stellar objects remains unclear, but a very high metallicity for the atmospheres of companions generated by core accretion as possibly found by \citet{Zhang2023} for AF Lep b and \citet{Samland2017} for 51 Eri b is likely. In any case, it signals a systematic difference in their evolution. Further progress in the modelling is needed to explain this observation. 

As a final point, we notice that the faintness of the L-T transition for companions - with respect to free-floating objects - may contribute to the low yields of surveys such as the SPHERE infrared survey for exoplanets (SHINE \citealp{Vigan2021}) and the Gemini Planet Imager Exoplanet Survey (GPIES \citealp{2019AJ....158...13N}), which were optimised for T-planet detections. This should be taken into account when estimating the frequency of giant planets from these surveys.

\begin{acknowledgements}
This work has made use of data from the European Space Agency (ESA) mission {\it Gaia} (\url{https://www.cosmos.esa.int/gaia}), processed by the {\it Gaia} Data Processing and Analysis Consortium (DPAC, \url{https://www.cosmos.esa.int/web/gaia/dpac/consortium}). Funding for the DPAC has been provided by national institutions, in particular, the institutions participating in the {\it Gaia} Multilateral Agreement.
This research has made use of the SIMBAD database, operated at CDS, Strasbourg, France. 
D.M., R.G., and S.D. acknowledge the PRIN-INAF 2019 'Planetary systems at young ages (PLATEA)' and 
ASI-INAF agreement n.2018-16-HH.0. A.Z. acknowledges support from ANID -- Millennium Science Initiative Program -- Center Code NCN2021\_080. S.M.\ is supported by the Royal Society as a Royal Society University Research  Fellowship (URF-R1-221669).
SPHERE is an instrument designed and built by a consortium consisting of IPAG (Grenoble, France), MPIA (Heidelberg, Germany), LAM (Marseille, France), LESIA (Paris, France), Laboratoire Lagrange (Nice, France), INAF-Osservatorio di Padova (Italy), Observatoire de Gen\`eve (Switzerland), ETH Zurich (Switzerland), NOVA (Netherlands), ONERA (France) and ASTRON (Netherlands), in collaboration with ESO. SPHERE was funded by ESO, with additional contributions from CNRS (France), MPIA (Germany), INAF (Italy), FINES (Switzerland) and NOVA (Netherlands). 
For the purpose of open access, the authors have applied a Creative Commons Attribution (CC BY) licence to any Author Accepted Manuscript version arising from this submission. 

\end{acknowledgements}

\bibliographystyle{aa} % style aa.bst
\bibliography{biblio} % your references Yourfile.bib

\begin{appendix}

\section{Photometry of sub-stellar objects}
\label{sec:photometry}

\begin{table*}
  \caption[]{Photometry for sub-stellar objects in the BPMG}
  \label{t:phot_BPMG}
  \scriptsize
  \setlength{\tabcolsep}{2pt}
\begin{tabular}{lcccccccccl}
  \hline
2MASS/Others	&	prob	&	par	&	$J$	&	$H$	&	$K$	&	M$_J$	&	M$_H$	&	M$_K$	&	$J-K$	&	Source	\\
		&	\%	&	mas	&	mag	&	mag	&	mag	&	mag	&	mag	&	mag	&	mag	&		\\
\hline
\multicolumn{11}{c}{Companions}\\
\hline
51 Eri b	b	&	8.9	&	33.44	&	19.779	$\pm$	0.200	&	19.489	$\pm$	0.210	&	18.489	$\pm$	0.14	&	17.40	$\pm$	0.20	&	17.11	$\pm$	0.21	&	16.11	$\pm$	0.14	&	1.29	$\pm$	0.24	&	\citet{Samland2017}	\\
AF Lep b	b	&	99.9	&	37.25	&	19.644	$\pm$	0.770	&	18.964	$\pm$	0.610	&	16.644	$\pm$	0.40	&	17.50	$\pm$	0.77	&	16.82	$\pm$	0.61	&	14.50	$\pm$	0.40	&	3.00	$\pm$	0.87	&	\citet{Mesa2023}\\
beta Pic b	b	&	99.9	&	51.44	&	14.250	$\pm$	0.200	&	13.223	$\pm$	0.200	&	12.473	$\pm$	0.05	&	12.81	$\pm$	0.20	&	11.78	$\pm$	0.20	&	11.03	$\pm$	0.05	&	1.78	$\pm$	0.21	&	\citet{Langlois2021}	\\
beta Pic c	c	&	99.9	&	51.44	&				&				&	14.343	$\pm$	0.10	&				&				&	12.90	$\pm$	0.10	&				&	\citet{Nowak2020}	\\
PZ Tel B	B	&	97.2	&	21.16	&	12.470	$\pm$	0.100	&	11.930	$\pm$	0.100	&	11.530	$\pm$	0.09	&	9.10	$\pm$	0.1	&	8.56	$\pm$	0.10	&	8.16	$\pm$	0.09	&	0.94	$\pm$	0.13	&	\citet{Maire2016}	\\
\hline
  \end{tabular}

$^a$ Photometric parallax
\normalsize
\end{table*}

\begin{table*}
  \caption[]{Photometry for sub-stellar objects in other young moving groups}
  \label{t:phot_ass}
  \scriptsize
  \setlength{\tabcolsep}{2pt}
  \begin{tabular}{lcccccccccl}
  \hline
2MASS/Others		&	memb	&	par	&	$J$	&	$H$&	$K$	&	$M_J$	&	$M_H$	&$M_K$	&	$J-K$	&	Source	\\
		&	\%	&	mas	&	mag	&	mag	&	mag	&	mag	&mag&mag	&	mag	&		\\
\hline
\multicolumn{9}{c}{Columba}\\
\hline
GSC 08047-00232	B	&	6.8	&	11.58	&	15.900	$\pm$	0.100	&	15.450	$\pm$	0.020	&	14.800	$\pm$	0.100	&	11.22	$\pm$	0.10	&	10.77	$\pm$	0.02	&	10.12	$\pm$	0.10	&	1.10	$\pm$	0.14	&	\citet{Patience2012}	\\
J00344300-4102266		&	58.9	&	21.54	&	15.707	$\pm$	0.067	&	14.807	$\pm$	0.065	&	14.084	$\pm$	0.056	&	12.37	$\pm$	0.07	&	11.47	$\pm$	0.07	&	10.75	$\pm$	0.06	&	1.62	$\pm$	0.09	&	\citet{Gagne2015}	\\
J03182597-3708118		&	98.6	&	13.16	&	13.366	$\pm$	0.024	&	12.757	$\pm$	0.021	&	12.404	$\pm$	0.023	&	8.96	$\pm$	0.02	&	8.35	$\pm$	0.02	&	8.00	$\pm$	0.02	&	0.96	$\pm$	0.03	&	\citet{Gagne2015}	\\
J03390160-2434059		&	74.9	&	17.08	&	10.898	$\pm$	0.030	&	10.335	$\pm$	0.035	&	9.968	$\pm$	0.025	&	7.06	$\pm$	0.03	&	6.50	$\pm$	0.04	&	6.13	$\pm$	0.03	&	0.93	$\pm$	0.04	&\citet{Gagne2015}	\\
J03550477-1032415		&	94.3	&	19.10	&	13.080	$\pm$	0.027	&	12.462	$\pm$	0.025	&	11.975	$\pm$	0.024	&	9.48	$\pm$	0.03	&	8.87	$\pm$	0.03	&	8.38	$\pm$	0.02	&	1.11	$\pm$	0.04	&	\citet{Shkolnik2017}	\\
\end{tabular}
  
$^a$ Photometric parallax
  \addtocounter{table}{-1}
\end{table*}

\begin{table*}
  \caption[]{Cont...}
  \scriptsize
  \setlength{\tabcolsep}{2pt}
  \begin{tabular}{lcccccccccl}
  \hline
2MASS/Others		&	memb	&	par	&	$J$	&	$H$&	$K$	&	$M_J$	&	$M_H$	&$M_K$	&	$J-K$	&	Source	\\
		&	\%	&	mas	&	mag	&	mag	&	mag	&	mag	&mag&mag	&	mag	&		\\
\hline
J22444835-6650032		&	99.7	&	22.82	&	11.030	$\pm$	0.026	&	10.413	$\pm$	0.021	&	10.143	$\pm$	0.024	&	7.82	$\pm$	0.03	&	7.20	$\pm$	0.02	&	6.93	$\pm$	0.02	&	0.89	$\pm$	0.04	&	\citet{Gagne2015}	\\
J22511530-6811216		&	99.9	&	21.01	&	12.101	$\pm$	0.024	&	11.591	$\pm$	0.027	&	11.239	$\pm$	0.024	&	8.71	$\pm$	0.02	&	8.20	$\pm$	0.03	&	7.85	$\pm$	0.02	&	0.86	$\pm$	0.03	&	\citet{Gagne2015}	\\
J23130558-6127077		&	98.2	&	21.65	&	10.932	$\pm$	0.021	&	10.373	$\pm$	0.025	&	10.050	$\pm$	0.023	&	7.61	$\pm$	0.02	&	7.05	$\pm$	0.03	&	6.73	$\pm$	0.02	&	0.88	$\pm$	0.03	&	\citet{Gagne2015}	\\
J23143092-5405313		&	99.9	&	22.91	&	11.501	$\pm$	0.026	&	10.976	$\pm$	0.022	&	10.636	$\pm$	0.021	&	8.30	$\pm$	0.03	&	7.78	$\pm$	0.02	&	7.44	$\pm$	0.02	&	0.86	$\pm$	0.03	&	\citet{Gagne2015}	\\
J23225240-6151114		&	99.9	&	23.56	&	11.534	$\pm$	0.023	&	10.963	$\pm$	0.022	&	10.654	$\pm$	0.019	&	8.40	$\pm$	0.02	&	7.82	$\pm$	0.02	&	7.52	$\pm$	0.02	&	0.88	$\pm$	0.03	&	\citet{Gagne2015}	\\
J23225299-6151275		&	96.7	&	23.24	&	15.545	$\pm$	0.062	&	14.535	$\pm$	0.063	&	13.860	$\pm$	0.043	&	12.38	$\pm$	0.06	&	11.37	$\pm$	0.06	&	10.69	$\pm$	0.04	&	1.69	$\pm$	0.08	&	\citet{Gagne2015}	\\
\hline
  \end{tabular}
  
$^a$ Photometric parallax\\
$^b$ Binary
  \addtocounter{table}{-1}
\normalsize
\end{table*}

\begin{table*}
  \caption[]{Cont...}
  \scriptsize
  \setlength{\tabcolsep}{2pt}
  \begin{tabular}{lcccccccccl}
  \hline
2MASS/Others		&	memb	&	par	&	$J$	&	$H$&	$K$	&	$M_J$	&	$M_H$	&$M_K$	&	$J-K$	&	Source	\\
		&	\%	&	mas	&	mag	&	mag	&	mag	&	mag	&mag&mag	&	mag	&		\\
 \hline
\multicolumn{9}{c}{Carina Near}\\
\hline
PSO J004.6359 +56.8370		&	90.8	&	47.00	&	16.220	$\pm$	0.020	&	16.240	$\pm$	0.020	&	16.130	$\pm$	0.020	&	14.58	$\pm$	0.02	&	14.60	$\pm$	0.02	&	14.49	$\pm$	0.02	&	0.09	$\pm$	0.03	&	\citet{Zhang2021}	\\
Koenigstuhl 1B		&	98.5	&	37.93	&	13.521	$\pm$	0.025	&	12.809	$\pm$	0.025	&	12.304	$\pm$	0.025	&	11.42	$\pm$	0.03	&	10.70	$\pm$	0.03	&	10.20	$\pm$	0.03	&	1.22	$\pm$	0.04	&	\citet{Hurt2023}	\\
WISE J003110.04+574936.3		&	97.5	&	71.00	&	14.950	$\pm$	0.040	&	13.780	$\pm$	0.040	&	13.215	$\pm$	0.029	&	14.21	$\pm$	0.04	&	13.04	$\pm$	0.04	&	12.47	$\pm$	0.03	&	1.74	$\pm$	0.05	&	\citet{Hurt2023}	\\
SIMP J013656.5 +093347.3		&	95.6	&	163.70	&	13.252	$\pm$	0.002	&	12.809	$\pm$	0.002	&	12.585	$\pm$	0.002	&	14.32	$\pm$	0.00	&	13.88	$\pm$	0.00	&	13.66	$\pm$	0.00	&	0.67	$\pm$	0.00	&	\citet{Zhang2021}	\\
J01033203+1935361		&	59.6	&	46.90	&	16.288	$\pm$	0.080	&	14.897	$\pm$	0.056	&	14.149	$\pm$	0.059	&	14.64	$\pm$	0.08	&	13.25	$\pm$	0.06	&	12.50	$\pm$	0.06	&	2.14	$\pm$	0.10	&	\citet{Schneider2017}	\\
\hline
  \end{tabular}

$^a$ Photometric parallax\\
$^b$ Binary
\normalsize
\end{table*}

\section{Masses of young sub-stellar objects with AMES models}
\label{sec:masses}

\begin{table*}
  \caption[]{Masses of sub-stellar objects with ages in the range 10-200 Myr derived using AMES models}
  \label{t:teff-r-mass}
  \begin{tabular}{lccccc}
  \hline
Name & Age & \teff(BC)    & \teff(remap) & $r$ & Mass \\
     & Myr &   K          & K            &   & \MJup \\
\hline
\multicolumn{6}{c}{Companions}\\
\hline
                51 Eri b&  21& 786$\pm$   8& 746$\pm$  10&  0.11$\pm$ 0.11&  4.11$\pm$ 0.30\\
                AF Lep b&  21& 950$\pm$  27& 940$\pm$   9&  0.29$\pm$ 0.20&  5.62$\pm$ 0.36\\
              beta Pic b&  21&1711$\pm$  63&1570$\pm$  32&  0.71$\pm$ 0.16& 11.83$\pm$ 0.91\\
        GSC08047-00232 B&  36&1988$\pm$  48&1975$\pm$  66&  0.49$\pm$ 0.30& 19.77$\pm$ 1.14\\
                HR8799 b&  36&1013$\pm$  35&1012$\pm$  18&  0.46$\pm$ 0.09&  7.24$\pm$ 0.58\\
\hline
  \end{tabular}
  \addtocounter{table}{-1}
\end{table*}

\begin{table*}
  \caption[]{Masses of sub-stellar objects with ages in the range 10-200 Myr derived using AMES models (cont...)}
  \begin{tabular}{lccccc}
  \hline
Name & Age & \teff(BC)    &\teff(remap) & $r$ & Mass \\
     & Myr &   K          & K            &   & \MJup \\
\hline
         J0045214+163445&  47&1888$\pm$  55&1777$\pm$  68&  0.86$\pm$ 0.11& 17.05$\pm$ 1.62\\
ULAS J004757.40+154641.4&  47&1101$\pm$  44&1092$\pm$  24&  0.17$\pm$ 0.02&  8.62$\pm$ 1.03\\
       J02074284+0000564&  47& 851$\pm$  10& 837$\pm$  19& -0.03$\pm$ 0.04&  6.27$\pm$ 0.82\\
       J02022917+2305141&  47&1358$\pm$  60&1302$\pm$  11&  0.46$\pm$ 0.09& 10.68$\pm$ 1.05\\
WISE J024124.73-365328.0&  47& &1289$\pm$   1&  0.59$\pm$ 0.26& 10.45$\pm$ 0.91\\
       J00303013-1450333&  47&1360$\pm$  60&1322$\pm$  32&  0.35$\pm$ 0.03& 10.73$\pm$ 1.13\\
\hline
  \end{tabular}
  \addtocounter{table}{-1}
\end{table*}

\begin{table*}
  \caption[]{Masses of sub-stellar objects with ages in the range 10-200 Myr derived using AMES models (cont...)}
  \begin{tabular}{lccccc}
  \hline
Name & Age & \teff(BC)    &\teff(remap) & $r$ & Mass \\
     & Myr &   K          & K            &   & \MJup \\
\hline
WISE J163645.56-074325.1& 137&1041$\pm$  38&1091$\pm$  23& -0.00$\pm$ 0.04& 13.68$\pm$ 1.02\\
WISE J174102.78-464225.5& 137&1442$\pm$  63&1451$\pm$  33&  0.57$\pm$ 0.03& 19.44$\pm$ 1.48\\
           PSO J306.0+16& 137&1729$\pm$  62&1791$\pm$  19&  0.76$\pm$ 0.08& 27.98$\pm$ 1.80\\
       J20490268-7456124& 137&1341$\pm$  60&1345$\pm$  26&  0.55$\pm$ 0.03& 17.67$\pm$ 1.15\\
           PSO J318.4+35& 137&1906$\pm$  54&2040$\pm$  48&  1.05$\pm$ 0.06& 36.66$\pm$ 3.10\\
       J21572060+8340575& 137&2068$\pm$  40&2238$\pm$  52&  1.10$\pm$ 0.00& 45.09$\pm$ 3.38\\
\hline
  \end{tabular}
  \addtocounter{table}{-1}
\end{table*}

\end{appendix}

\end{document}